\documentclass[aps,pra,twocolumn,groupedaddress,amsmath,amssymb,
aps]{revtex4-2}

\usepackage{graphicx}
\usepackage{dcolumn}
\usepackage{bm}
\usepackage{braket}
\usepackage{amsmath}

\begin{document}

\title{Non-commutative graphs based on finite-infinite system couplings: quantum error correction for a qubit coupled to a coherent field}


\author{G.G.~Amosov}
	\email{gramos@mi-ras.ru}
\affiliation{Steklov Mathematical Institute of Russian Academy of Sciences, Gubkina str. 8, Moscow 119991, Russia}
\author{A.S.~Mokeev}
	\email{alexandrmokeev@yandex.ru}
\affiliation{Steklov Mathematical Institute of Russian Academy of Sciences, Gubkina str. 8, Moscow 119991, Russia}
\author{A.N.~Pechen}
	\email{apechen@gmail.com}
\homepage[]{http://www.mathnet.ru/eng/person17991}
\affiliation{Department of Mathematical Methods for Quantum Technologies, Steklov Mathematical Institute of Russian Academy of Sciences, Gubkina str. 8, Moscow 119991, Russia}
\affiliation{National University of Science and Technology "MISIS", Leninsky prosp. 6, Moscow 119991, Russia}

\date{\today}

\begin{abstract}
Quantum error correction plays a key role for quantum information transmission and quantum computing. In this work, we develop and apply the theory of non-commutative operator graphs to study error correction in the case of a finite-dimensional quantum system coupled to an infinite dimensional system. We consider as an explicit example a qubit coupled via the Jaynes-Cummings Hamiltonian with a bosonic coherent field. We extend the theory of non-commutative graphs to this situation and construct, using the Gazeau-Klauder coherent states, the corresponding non-commutative graph. As the result, we find the quantum anticlique, which is the projector on the error correcting subspace, and analyze it as a function of the frequencies of the qubit and the bosonic field. The general treatment is also applied to the analysis of the error correcting subspace for certain experimental values of the parameters of the Jaynes-Cummings Hamiltonian. The proposed scheme can be applied to any system that possess the same decomposition of spectrum of the Hamiltonian into a direct sum as in JC model, where eigenenergies in the two direct summands form strictly increasing sequences.
\end{abstract}

\maketitle

\section{Introduction}
	
Quantum error correcting codes (or, in other terminology, quantum anticliques), introduced theoretically in the pioneering papers~\cite{shor, Ekert1996, Knill1997}, and implemented experimentally, e.g., in~\cite{Cory1998}, play an important role in quantum information theory~\cite{ManchiniWinterBook}. Analog error correction for continuous variables, such as position and momentum, was considered~\cite{Lloyd1998}, symmetry breaking in open quantum systems for photonic cat qubits~\cite{Lieu2020}, etc.  In particular, bosonic codes which use encoding of information in states of bosonis field are of high interest~\cite{Cochrane1999,Gottesman2001,Albert2018,TerhalQST2020,Albert2020}.  
	
Error correction theory studies the possibility of encoding information in quantum states in a way to allow zero-error decoding in the presence of a given fixed acceptable set of errors. Mathematically each error is described by some completely positive map acting on the set of states of the quantum system. In the general setting~\cite{knill}, for any given set of errors it is possible to define a unique non-commutative operator graph \cite{Duan} such that the knowledge of this graph allows to define all error correcting codes for this set of errors. The correspondence between sets of errors and non-commutative operator graphs is not one-to-one, but in the case of separable Hilbert space each non-commutative operator graph describes codes for some set of errors~\cite{Duan2, shirokov, yashin}. As was established in the finite-dimensional \cite {Duan2, shirokov} and infinite-dimensional \cite {yashin} cases, the noncommutative graph describing errors in information transmission is always generated by some positive operator-valued measure.
	
Various models of error correction were analyzed using the approach based on the use of non-commutative graphs. It was applied for quantum error correction for the models of coupled finite-dimensional systems~\cite{AM1,AM2,AM3}, and for coupled infinite-dimensional system~\cite{AMP}. In \cite{AMP}, the non-commutative graph generated by the dynamics of a bipartite bosonic quantum system in an infinite dimensional Hilbert space was defined. The graph consists of orbits driven by the unitary group which is the solution of the Schr\"oedinger equation for a two interacting bosonic oscillators. In this framework possible error correcting codes are given by coherent states in the bosonic Fock space. In all these cases, it was possible to find quantum anticliques, which are projectors onto error correcting subspaces. 

In this work, we extend the theory of operator graphs to the case when one system is finite dimensional while another is infinite dimensional. Currently, several quantum error-correction \cite{Kyaw1,Wu} and entanglement protection \cite{Fasihi} techniques was inroduced for systems of such structure. Explicitly, we consider the situation where the information is encoded in a joint state of a qubit (a two-level quantum system) coupled via the Jaynes-Cummings Hamiltonian~\cite{JayCum} with a bosonic oscillator or bosonic coherent field. The Jaynes-Cummings Hamiltonian is the key model for the various theoretical and experimental works in quantum optics and for studying the interactions between light and matter, see, e.g.~\cite{Rempe1987,Forn,Kockum,Casanova, Irish,Huang,Yoshihara,Langford,Cirac1993,Cirac1994,Haroche,Blais2004,Larson2007,Saffman2010}, including in strong~\cite{Cirac1993}, ultra-strong~\cite{Forn,Kockum,Huang} and deep strong coupling regimes~\cite{Forn,Kockum,Casanova,Huang,Yoshihara,Langford}.

We develop for this model the theory of non-commutative operator graphs and apply it to find the corresponding quantum anticlique. The construction is based upon Gazeau--Klauder coherent states~\cite{GeaKla}. With this setting, we explicitly find the error correcting subspace for any values of the parameters of the Jaynes-Cummings model. 
	
The structure of the paper is the following. In section 2, we discuss the problem of finding the existence of an error correcting procedure for a quantum channel, and also describe quantum channels corresponding to operator graphs of the class which includes the non-commutative graphs later constructed in section 5. In section 3, the Jaynes-Cummings model is discussed. Section 4 describes the construction of Gazeau-Klauder coherent states. In section 5, using the Gazeau-Klauder coherent states for the Jaynes-Cummings model, we construct the non-commutative operator graphs that have quantum error correcting codes, and find quantum anticlique and error correction subspaces.
	
\section{Quantum channels and non-commutative graphs}
	
Consider encoding information in states of a quantum system with Hilbert space $\cal H$. The (convex) set  $\mathfrak {S}({\cal H})$ of quantum states is the set of positive unit trace operators in $\cal H$. Errors which can occur under information transmission can be described by a quantum channel $\Phi  :\mathfrak {S}({\cal H})\to \mathfrak {S}({\cal H})$ which is a completely positive trace preserving (CPTP) map. As any CPTP map, it possesses  the Kraus operator-sum representation (Kraus OSR) \cite {Hol}
\begin{equation}\label{count1}
\Phi (\rho )=\sum \limits _{k\in K}V_k\rho V_k^*,\quad \rho \in \mathfrak {S}(\mathcal{H}).
\end{equation}
The Kraus operators $\left\lbrace V_k,\ k\in K\right\rbrace $ are parametrized by some set $K$. 
They should satisfy the property
\begin{equation}\label{count2}
\sum \limits _{k\in K}V_k^*V_k=\mathbb I.
\end{equation}
to preserve trace of density matrix. In infinite dimensional spaces $K$ is not necessary countable and the sum in (\ref {count1}) -- (\ref {count2}) can be replaced by an integral (see e.g. \cite{SI}). Nevertheless for any channel $\Phi $ there exists a countable set $K$ parametrizing Kraus operators such that (\ref {count1}) holds true. Note that the Kraus OSR of a quantum channel is non-unique. The same quantum channel also has a Kraus OSR with operators $\tilde V_j=\sum\limits_i U_{ji}V_i$, $j=1,\dots,m$ ($m\ge |K|$), where $U^+U=\mathbb I$ (unitary matrix). 
	
The linear space, $\mathcal{V}$, plays an important role in the theory of optimal coding
\begin{equation}\label{gr}
\mathcal {V}=\overline {span} \left\lbrace V_k^*V_j,\ k,j\in K\right\rbrace .
\end{equation}
The linear space $\mathcal V$ does not depend on the choice of the operators $\{V_i\}$ used for Kraus OSR of a given quantum channel; despite the non-uniqueness of the Kraus OSR, it is unique for a given quantum channel. Notice that in the finite dimensional case there is not need to take the closure in~(\ref {gr}). For the infinite dimensional case see~\cite{AMP,yashin}. 
	
The linear space  $\mathcal V$ has the properties of a non-commutative graph. Such objects were introduced in~\cite{ChoiEffros} as operator systems and recently redefined as non-commutative graphs in quantum information theory~\cite{Duan}. 
A {\it non-commutative graph} is a linear subspace $\mathcal V$ of bounded operators in a Hilbert space ${\cal H}$ possessing the properties
\begin{itemize}
\item ${\bf V}\in \mathcal V$ implies that ${\bf V}^*\in \mathcal V$;
\item ${\bf I}\in {\mathcal V}$
\end{itemize}
	
The famous Knill--Laflamme condition\cite{Knill1997,knill} claims that a zero error transmission via some channel $\Phi $ is possible iff for some orthogonal projector $P$ for all $A\in \mathcal{V}$ holds $PAP=\alpha(A)P$, where $\alpha(A)\in \mathbb{C}$. Here $P$ is the projector on the subspace generated by error correction code~\cite{Knill1997}. The optimal code belongs to the subspace ${\cal H}_P=P{\cal H}$.  The dimension of the subspace ${\cal H}_P$ is the maximal amount of quantum information that could be transmitted via $\Phi$ with zero error. An orthogonal projection $P$ such that $\dim\left(  P {\cal H}\right) \ge 2$ is a {\it quantum anticlique} for a non-commutative graph $\mathcal V$ if it satisfies:
\begin{equation}\label {anti}
\dim P{\mathcal V} P=1.
\end{equation}
	
The most natural quantum channel is given by the projection measurement
\begin{equation}\label{channel}
\Phi _{\mathcal P}(\rho )=\sum \limits _{k\in K}P_k\rho P_k,\quad \rho \in \mathfrak {S}({\cal H}),
\end{equation}
where $\mathcal {P}=(P_k)$ is the orthogonal resolution of identity
$$
\sum \limits _{k\in K}P_k=\mathbb I.
$$
For the channel (\ref{channel}), the non-commutative graph (\ref{gr}) is
$$
\mathcal {V}=\overline {span} \left\lbrace P_k, k\in K \right\rbrace.
$$
In the finite dimensional case, it is enough to consider only discrete sets $K$, while for the infinite dimensional case $P_k=E(B_k)$ are generated by some projection valued measure on the real line, where $B_k\subset {\mathbb R}$ are some Borel sets possessing the property $\cup_{k\in K}B_k={\mathbb R}$. In this case, different choices of $B_k$'s produce different projectional measurements and a reachable set of admissible errors.
	
Suppose that some unitary group ${\mathcal U}=\{ U_t=e^{-itG},\ t\ge 0\}$ acts in the Hilbert space $H$. Then the possible set of errors can be extended to all  possible projection measurements
\begin{equation}\label{e*}
\Phi ^t(\rho )=\sum \limits _{k\in K}U_tP_kU_t^*\rho U_tP_kU_t^*,\ t\in {\mathbb R},
\end{equation}
For the goal  of constructing a quantum anticlique allowing to correct errors of the form (\ref {e*}) for any fixed $t$, it is natural to define the non-commutative graph corresponding to all these errors as follows
$$
{\mathcal V}=\overline {span}\{U_tP_kU_t^*,\ t\in {\mathbb R},\ k\in K\}.
$$
Our interpretation of such a graph is that the quantum system distorts the transmitted information by a set of time-dependent errors.
	
Based upon this interpretation, suppose that there is a set of orthogonal projections $\left\lbrace P_{\alpha },\ \alpha \in \mathfrak {A}\right\rbrace $ parameterized by some set $\mathfrak {A}$, and the operator space is generated by orbits of some unitary group $\mathcal U$ as follows
\begin{equation}\label{first}
{\mathcal V}=\overline {span}\{U_tP_{\alpha }U_t^*,\ t\in {\mathbb R},\ \alpha \in \mathfrak {A} \}.
\end{equation}
It is known \cite {Duan2, yashin} that (\ref {first}) is a non-commutative operator graph corresponding to some channel iff $\mathbb I\in \mathcal V$. We shall construct the explicit example of graph for the Jaynes-Cummings model. Moreover, we shall show that there exists an anticlique for this graph.
	
\section{Jaynes-Cummings model}
	
\begin{figure}[t]
\centering
\includegraphics[width=1\linewidth]{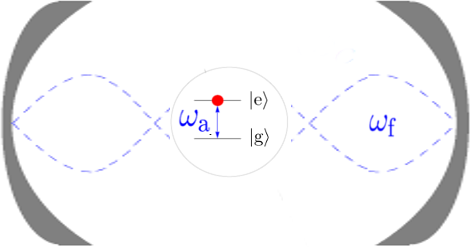}
\caption{\label{figureJC} Jaynes-Cummings model of a qubit interacting with bosonic reservoir.}
\end{figure}

We consider a two-level quantum system (qubit) coupled to a coherent field. Hilbert space of the qubit is ${\cal H}_{\rm s}=\mathbb C^2$. Ground and excited basis states of the qubit are denoted as $\{\ket{g},\ket{e}\}$. Hilbert space of the coherent field is ${\cal H}_{\rm f}=L^2(\mathbb R)=\{f:\mathbb R\to\mathbb C\,|\, \int_\mathbb R|f(x)|^2 dx<\infty\}$. Fock states of the field are denoted as  $\{\ket{k}, k\in \mathbb{N}_0\}$. We use the set of natural numbers including zero $ \mathbb {N}_0 = {0} \cup \mathbb {N}$ to enumerate the states. 
The qubit and the field are assumed to be coupled via the Jaynes-Cummings Hamiltonian acting in ${\cal H}={\cal H}_{\rm s}\otimes {\cal H}_{\rm f}$ 
\begin{equation}\label{HAM}
H=\omega_f a^{+}a^{-}+\frac{\omega_s}{2} \sigma_z + \frac{\kappa}{2}(\sigma^{-} a^{+}+\sigma^{+}a^{-}),
\end{equation}
Here $\omega_s, \omega_f \in \mathbb{R_+}$ are the frequencies of the qubit and the field, respectively, $\kappa\ge 0$ is the coupling constant, $\sigma_z$ is the Pauli matrix, $\sigma^{+},\sigma^{-}$ are the rising and lowering operators of the qubit and the $a^{+},a^{-}$ are the creation and annihilation operators of the field. The detuning parameter is $\Delta=\omega_f-\omega_s$. We use the normalization of the physical units such that $\hbar=1$. Denote the basis in ${\cal H}$ as $\ket{q}\otimes\ket{p}=\ket{q,p}$, where the first number $q\in \mathbb{N}_0$ denotes the coherent state and the second number $p\in \{e,g\}$ denotes the qubit state. The schematic picture of the Jaynes-Cumming model states interaction is provided in Fig.~\ref{figureJC}.
	
The Schr\"oedinger equation with the Jaynes-Cummings Hamiltonian has an exact solution. The Hamiltonian has the following eigenstates
\begin{eqnarray*}\label{dressedstates}
		&& \ket{0,g},\\
		&&\ket{n,+}=\cos \left(\frac{\theta_n}{2}\right)\ket{n-1,e}+\sin \left(\frac{\theta_n}{2}\right)\ket{n,g},\\
		&&\ket{n,-}=\sin \left(\frac{\theta_n}{2}\right)\ket{n-1,e}-\cos \left(\frac{\theta_n}{2}\right)\ket{n,g},
\end{eqnarray*}
where $\theta_n=\tan^{-1}(\kappa\sqrt{n}/\Delta)$ and $n\in \mathbb{N}$, for the non-resonant case $\Delta\neq 0$. For the resonant case $\Delta = 0$ the eigenstates are
\begin{eqnarray*}\label{dressedstates}
		&& \ket{0,g},\\
		&&\ket{n,+}=\ket{n-1,e}+\ket{n,g},\\
		&&\ket{n,-}=\ket{n,g}-\ket{n-1,e}.
\end{eqnarray*}
In both cases the corresponding eigenenergies are
\begin{eqnarray*}
		E_{0,g}&=& \frac{\omega_f+\Delta}{2} \\
		E_{n,\pm}&=&\omega_f\left(n-\frac{1}{2}\right)\pm \frac{1}{2}\sqrt{\Delta^2 + \kappa^2 n},\quad  n\in \mathbb{N}.
\end{eqnarray*}
	
Below we follow closely to the method provided in \cite{Hussin}, where a new class  of coherent states was constructed for the Jaynes-Cummings model with strictly increasing sequences of the eigenenergies $E_{n,+}$ and $E_{n,-}$. Our goal is to divide the space $\cal H$ into three direct summands, two of which are generated by eigenstates corresponding to strictly increasing sequences of eigenenergies and one is finite dimensional. The sequence 
\begin{equation}\label {J}
		J_k=E_{k+1,+}, \quad k \in \mathbb{N}_0
\end{equation} 
is known to be strictly increasing. On the other hand, the sequence $S_0=E_{0,g},S_k=E_{k,-},\ k\in\mathbb{N}$ may have degenerate levels $S_{k_1}=S_{k_2},\ k_1\neq k_2$. We want to keep only a strictly increasing tail of the sequence $S_k$. Let us show that there exists $M_0 \in \mathbb{N}$ such that for all  $l_2>l_1\ge M_0$ one gets $S_{l_2}> S_{l_1}$. It is equivalent to
\begin{widetext}
\begin{equation}
	S_{n+1}-S_n
	=\omega_f-\frac{1}{2}\left(\sqrt{\Delta^2+\kappa^2(n+1)}-\sqrt{\Delta^2+\kappa^2n}\right)>0,
	\quad \forall n\ge M_0.
\end{equation}
\end{widetext}
From this one gets that $M_0$ is the minimal integer solution of the inequality
\begin{equation}\label{K0}
		\left(\sqrt{\Delta^2+\kappa^2(M_0+1)}+\sqrt{\Delta^2+\kappa^2M_0}\right)^{-1}<\frac{2\omega_f}{\kappa^2}.
\end{equation}
Thus, the sequence
\begin{equation}\label{S}
		S_k=E_{k,-},\quad k\ge M_0,
\end{equation}
becomes strictly increasing. 
	
Let us fix any number $K_0\in\mathbb N$, $K_0\ge M_0$. Then the sequence $ S_k=E_{k,-}$, $k\ge K_0$ will also be strictly increasing and we can  separate the pieces where we assured to have strictly increasing eigenenergies. This allows to represent the Hilbert space $\cal H$ as the direct sum ${\cal H}={\cal H}_1\oplus {\cal H}_2\oplus {\cal H}_3$, where
\begin{eqnarray}
		{\cal H}_1&=&span\{\ket{n,+},\ n\in \mathbb{N}\},\\
		{\cal H}_2&=&span\{\ket{n,-},\ n\ge K_0\},\\
		{\cal H}_3&=&span\{\ket{g,0}\}\cup\{\ket{n,-},\ 1\le n < K_0\},
\end{eqnarray}
The subspace ${\cal H}_3$ later will be shown to be the error correcting subspace for this system. The error dynamics will be shown to interchange states in ${\cal H}_1$ and ${\cal H}_2$, while keeping states in $\mathcal{H}_3$ unchanged. 
	
In Section 5, following the ideas of~\cite{Hussin} we will define the Gazeau-Klauder coherent states in ${\cal H}_1$ and ${\cal H}_2$.
	
\section{Gazeau-Klauder coherent states}
	
Here we introduce the construction of Gazeau-Klauder coherent states \cite{GeaKla}. Let us consider an infinite dimensional Hilbert space ${\cal H}$ with the basis $\ket{k},k\in \mathbb{N}_0$ and a self-ajoint operator $G$ which is diagonal in this basis. In \cite{GeaKla} Gazeau and Klauder defined the generalized coherent states corresponding to the operator $G$ as a two-parameter system of vectors $\{\ket{x,y}, x\in \mathbb{R_+},\ y\in \mathbb{R} \}\subset \mathcal{H}$ with the following properties
\begin{enumerate}\label{gcs}
		\item \textit{Continuity:}$\ (x,y)\rightarrow(x_0,y_0)\Rightarrow\ket{x,y}\rightarrow \ket{x_0,y_0}$.
		\item \textit{Resolution of identity:} $\int \ket{x,y}\bra{x,y} d\nu(x,y)=\mathbb I_{\cal H}$.
		\item \textit{Temporal stability:} $e^{-itG}\ket{x,y}=\ket{x,y+\omega t}$.
		\item \textit{Action identity:} $\bra{x,y} G\ket{x,y}=\omega x$.
\end{enumerate}
for some real constant $\omega$ and some measure $\nu$. 
	
Consider the set of eigenvalues $h_k=\bra{k}G\ket{k}$ for the operator $G$. In the case $h_0=0$ and strictly increasing $h_k$, Gazeau and Klauder gave the explicit construction for  the system of coherent states. If $h_0> 0$ and the sequence $h_k$ is strictly increasing, their construction describes the set of vectors that satisfy the first two properties and the following version of the time stability condition:
$$
e^{-itG}\ket{x,y}\bra{x,y}e^{itG}=\ket{x,y+\omega t}\bra{x,y+\omega t}.
$$
	
The main property of generalized coherent states is that they form the resolution of identity. It was shown that the measure $\nu $ has the form
$$
d\nu (x,y)=\tau (x)dxdy,
$$
where $\tau (x)$ is some probability distribution density on the half-axis.
	
Below we give an explicit description of this construction. Consider a sequences of weights
$$
c_k>0,\quad k \in \mathbb{N}_0,
$$
with the convergence condition
\begin{equation}\label{conv}
\limsup_{k\rightarrow \infty} \sqrt [k]{c_k}=R >0.
\end{equation}
Suppose that these weights are the moments of probability distributions with  density $\rho(x)> 0$ on the interval $[0,R)$,
$$
c_k=\int\limits_0^R \rho(x) x^k dx < +\infty,\quad  k \in \mathbb{N}_0.
$$
We also need the normalization factor and the density defined by the formulae
\begin{eqnarray}\label{ser}
N^2(x)&=&\sum_{k=0}^{\infty}\frac{x^k}{c_k},\quad  0\le x< R,\\
\tau(x)&=&N^2(x)\rho (x).\nonumber
\end{eqnarray}
The radius of convergence in (\ref {ser}) is equal to $R$ by the property (\ref{conv}).  Now the Gazeau--Klauder coherent states are defined as follows
\begin{equation}\label{GKS}
\ket{x,y}=\frac{1}{N(x)}\sum_{k=0}^{\infty}\frac{x^{k/2}e^{-ih_{k}y}}{\sqrt{c_k}}\ket{k}
\end{equation}
We suppose the constant $\omega$ is equal to one.
Following the definition in \cite{GeaKla}, for $f:\mathbb{R}\rightarrow B(\mathcal{H})$, where $B(\mathcal{H})$ is the set of bounded linear operators on $\mathcal{H}$, we introduce its integration as the weak-limit of averages of weak integrals, 
$$
I(f)=\int\limits_{-\infty}^{+\infty}f(y)d\mu(y)=\lim_{R \rightarrow +\infty} \frac{1}{2R}\int\limits_{-R}^{R}f(y)dy.
$$
Note that if such integral converges for some $f$ with the image lying in the weakly-closed subspace $Im(f)\subset\mathcal{W}\subset B(H)$, then the integral also lies in this subspace, $I(f)\in\mathcal{W}$. The resolution of identity property for coherent states results in
\begin{equation}\label{resol}
\int\limits_{0}^{R}\int\limits_{-\infty}^{+\infty}\ket{x,y}\bra{x,y}\tau(x)dxd\mu(y)=\mathbb I_{\cal H}. 
\end{equation}

\section{Graphs generated by Gezeau-Klauder coherent states}

Now we are able to define systems of Gazeau--Klauder coherent states \cite{GeaKla} in the Hilbert spaces ${\cal H}_1$ and ${\cal H}_2$. Take two sequences of weights
$$
c_k^{(j)}>0,\quad d \in \mathbb{N}_0,\   j=1,2,
$$
with the same convergence condition
$$
\limsup_{k\rightarrow \infty} \sqrt[k]{c_k^{(j)}}=R >0.
$$
Suppose that the weights have the corresponding probability densities $\rho_1(x),\rho_2(x)> 0$ on the interval $[0,R]$ such that
$$
c_k^{(j)}=\int\limits_0^R \rho_j(x) x^k dx < +\infty,\quad k \in \mathbb{N}_0,\   j=1,2.
$$
Then, the normalization factors and the densities for measures defining resolutions of identity are given by the formulae
\begin{eqnarray*}
N_{j}^2(x)&=&\sum_{k=0}^{\infty}\frac{x^k}{c_k^{(j)}}, \quad 0\le x <R,\\
		\tau_{j}(x)&=&N_{j}^2(x)\rho_j(x),\quad j=1,2.
\end{eqnarray*}
Consider the Gazeau--Klauder coherent states
\begin{eqnarray*}
		\ket{J,x,y}&=&\frac{1}{N_1(x)}\sum_{k=0}^{\infty}\frac{x^{k/2}e^{-iJ_ky}}{\sqrt{c_k^{(1)}}}\ket{k+1,+},\\
		\ket{S,x,y}&=&\frac{1}{N_2(x)}\sum_{k=0}^{\infty}\frac{x^{k/2}e^{-iS_{k+K_0}y}}{\sqrt{c_k^{(2)}}}\ket{k+K_0,-},
\end{eqnarray*}
where the strictly increasing sequences of eigenenergies are given by (\ref {J}) and (\ref {S}) respectively.
	
Since the Gazeau--Klauder coherent states section \ref {gcs} form the resolution of identity, we get for the projections $P_{{\cal H}_1},P_{{\cal H}_2}$ on the subspaces ${\cal H}_1, {\cal H}_2$ that
\begin{equation}\label{resol1}
		\int\limits_{0}^{R}\int\limits_{-\infty}^{+\infty}\ket{J,x,y}\bra{J,x,y}\tau_1(x)dx d\mu(y)=P_{{\cal H}_1}, 
\end{equation}
\begin{equation}\label{resol2}
		\int\limits_{0}^{R}\int\limits_{-\infty}^{+\infty}\ket{S,x,y}\bra{S,x,y}\tau_2(x)dxd\mu(y)=P_{{\cal H}_2}.
\end{equation}
	
Consider the unitary group ${\mathcal U}=\{ U_t=e^{-it{H}},\ t\in {\mathbb R}\}$, where the Hamiltonian $H$ is determined by~(\ref {HAM}). Systems $\ket{J,x,y}, \ket{S,x,y}$ satisfy the temporal stability property~(\ref {gcs}) with respect to $\mathcal U$,
\begin{equation}\label{ts1}
		U_t\ket{J,x,y}\bra{J,x,y}U_t^{*}=\ket{J,x,y+t}\bra{J,x,y+t}
\end{equation}
\begin{equation}\label{ts2}
		U_t\ket{S,x,y}\bra{S,x,y}U_t^{*}=\ket{S,x,y+t}\bra{S,x,y+t}
\end{equation}
	
Consider the two families of orthogonal projections
$$
P_x^1=\ket{J,x,0}\bra{J,x,0},\quad  P_x^2=\ket{S,x,0}\bra{S,x,0},
$$
for $x\in [0,R]$. The projections $P_x^1$, $P_x^2$ and $P_x^3\equiv P_{{\cal H}_3}$ are pairwise orthogonal for any fixed value of $x\in [0,R]$.
	
{\bf Theorem 1.} The subspace
$$
	\mathcal{V}=\overline{span}\lbrace U_tP_x^jU_t^{*},\ t\in \mathbb{R}, \ x \in [0,R] ,\ j\in \{1,2,3\}\rbrace
$$
is a non-commutative operator graph with the anticlique $P_{{\cal H}_3}.$
	
\noindent {\bf Proof.} Consider the operator
$$
Q_x=P_x^1+\frac {\tau _2(x)}{\tau _1(x)}P_x^2+\frac {1}{\tau _1(x)}P_{{\cal H}_3}
$$
It follows from (\ref {ts1}) and (\ref {ts2}) that
\begin{eqnarray*}
	U_tQ_xU_t^{*}\\
	=\ket{J,x,t}\bra{J,x,t}+\frac{\tau_2(x)}{\tau_1(x)}\ket{S,x,t}\bra{S,x,t}+ \frac{1}{\tau_1(x)}P_{{\cal H}_3}.
\end{eqnarray*}
Then, (\ref{resol1}) and (\ref {resol2}) result in
\begin{widetext}
\begin{equation}
				\int\limits_{0}^{R}\int\limits_{-\infty}^{+\infty}\tau_1(x)\left(\ket{J,x,t}\bra{J,x,t}
			+ \frac{1}{\tau_1(x)}\mathbb I_{{\cal H}_2}+ \frac{\tau_2(x)}{\tau_1(x)}\ket{S,x,t}\bra{S,x,t}\right)
			dxd\mu(t)=\mathbb I_{\cal H}\in {\mathcal V}.
\end{equation}
\end{widetext}
Since $K_0$ is given by the rule (\ref{nontriv}) the dimension of ${\cal H}_3$ is at least 2. From the equalities
\begin{eqnarray*}
		P_{{\cal H}_3}\ket{J,x,t}\bra{J,x,t}P_{{\cal H}_3}&=&0\\
		P_{{\cal H}_3}\ket{S,x,t}\bra{J,x,t}P_{{\cal H}_3}&=&0
\end{eqnarray*}
we obtain that $P_{{\cal H}_3}$ is an anticlique. $\Box$
	
\section{The error correcting subspace}
As Theorem~1 states, the subspace ${\cal H}_3$ is the error correcting subspace. In some cases the number is $M_0=1$, so in this case for $K_0=M_0$ the error correcting subspace would be empty (since its dimension is $K_0-1$). However, in our construction one can take any natural $K_0\ge M_0$. To satisfy this condition for our coding procedure, that is the dimension of error correcting subspace is greater or equal to two, we should take any
	
\begin{equation}\label{nontriv}
		K_0\ge K_0^*=\max\{3,M_0\}.
\end{equation}
	
\begin{figure}[h]
		\includegraphics[width=1\linewidth]{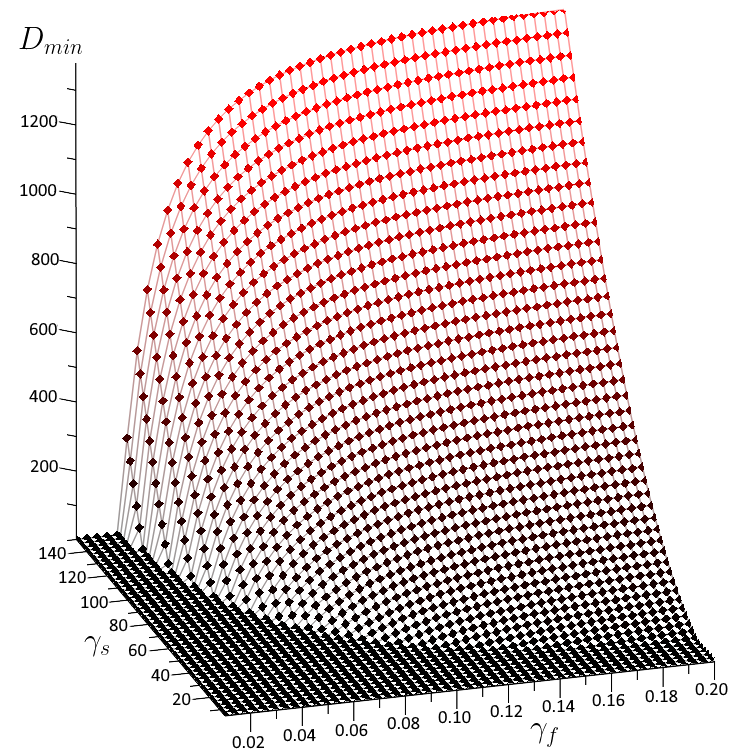}
		\caption{\label{figureBeh} Behavior of the minimal dimension of the error correcting subspace vs coupling rates of the Jaynes-Cummings Hamiltonian.}
\end{figure}

\begin{figure}[h]
		\includegraphics[width=1\linewidth]{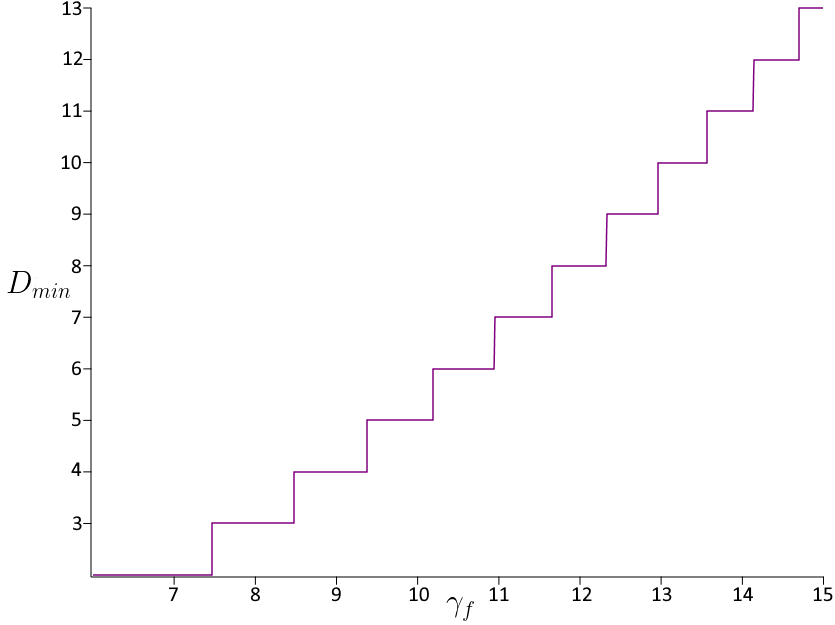}
		\caption{\label{figureRes} Behavior of the minimal dimension of the error correcting subspace for the resonant case $\Delta=0$ vs coupling rate of the field.}
\end{figure}

To analyze the minimal dimension of the error correcting subspace for various parameters of the Jaynes-Cummings Hamiltonian, consider the coupling rates $\gamma_f=\kappa/ \omega_f$ and $\gamma_s=\kappa/ \omega_s$. In terms of these quantities the inequality~(\ref{K0}) takes the following form
\begin{widetext}
\begin{equation} \label{mindim}
	 \sqrt{(\gamma_f^{-1}-\gamma_s^{-1})^2+M_0+1}+\sqrt{(\gamma_f^{-1}-\gamma_s^{-1})^2+M_0} >\frac{\gamma_f}{2}.
\end{equation}
\end{widetext}
Now it is evident that the key quantity defining the dimension for the resonant case (when  $\Delta=\omega_f-\omega_s=0$ and hence $\gamma_f=\gamma_s$) is the coupling rate $\gamma_f$. For having $M_0$ equal or larger than $4$ or greater, we need the $\gamma_f$ to be at least $2(2+\sqrt{3})$. For non-resonant case for fixed $\gamma_f$ decreasing the value $\gamma_s$ will decrease $M_0$. Figure \ref{figureBeh} shows the behavior of the minimal possible dimension $D_{\rm min}=K_0^*-1$ of the error correcting subspace ${\cal H}_3$ vs coupling rates $\gamma_s$ and $\gamma_f$. The figure clearly shows that the behavior is non-symmetric with respect to $\gamma_s$ and $\gamma_f$, as is also evident from inequality~(\ref{mindim}). The resonant case is the extremal case in the inequality~(\ref{mindim}), what means if $M_0$ satisfies it for some $\gamma_f$, then for the same parameter $\gamma_f$ the number $M_0$ also solves the inequality in the resonant case. Figure~\ref{figureRes} shows the dependence of $D_{\rm min}$ on the coupling rate $\gamma_f$ for equal frequencies.
   
The Jaynes-Cummings Hamiltonian is used in various theoretical and experimental analysis in quantum optics, cavity QED, e.g.~\cite{Rempe1987,Forn,Kockum,Casanova, Irish,Huang,Yoshihara, Langford,Cirac1993,Cirac1994,Haroche,Blais2004,Larson2007,Saffman2010}, including in strong~\cite{Cirac1993}, ultra-strong~\cite{Forn,Kockum,Huang} and deep strong coupling regimes~\cite{Forn,Kockum,Casanova,Huang,Yoshihara,Langford}. We take for example of the weak coupling regime the parameters used in experiments performed in the group of Serge Haroche~\cite{Haroche}. In the setup of that experiments the cavity is designed to have the frequency equal to the frequency of the atom, i.e. $\Delta = 0$.   The approximate parameters for the experiment are $\kappa=2\pi \cdot 47$ KHz and $\omega_f=\omega_s=2\pi\cdot 51.1$ GHz. For this case the inequality~(\ref{K0}) becomes 
\begin{equation} \label{mindim1}
	\sqrt{M_0+1}+\sqrt{M_0}>\frac{\kappa}{2\omega_f}=\frac{47}{2\cdot51.1\cdot10^6} \approx 0.46 \cdot 10^{-6}
\end{equation}
The minimal natural solution of this inequality is $M_0=1$, so $K_0=3$ and the minimal dimension is $D_{\rm min}=K_0-1=2$. This minimal two-dimensional error correcting subspace is spanned by the two vectors $\ket{g,0}$ and $\ket{1,-}=\ket{1,g}-\ket{0,e}$. We remark that this is the error correcting subspace of minimal dimension. One could choose arbitrary large $K_0$ and the corresponding non-commutative operator graph will have the error correcting subspace $H_3$ of dimension $K_0-1$.
	
Jaynes-Cummings model is derived from Rabi model via the rotating wave approximation (RWA). This approximation is typically valid for $\gamma_f<0.1$ and $\omega_f\approx \omega_s$. In this case the minimal dimension is $D_{\rm min}=2$.  As it can be seen from Fig.~\ref{figureBeh} and  Fig.~\ref{figureRes}, to have $D_{\rm min}>2$ one has to consider values of $\gamma_f$ in the range of deep strong coupling regime~\cite{Forn,Kockum,Casanova}. This regime, as well as less intense ultrastrong regime, is of interest now. In these regimes the Rabi model is non-integrable and investigating these regimes motivates describing eigen-energies approximations for this model~\cite{Casanova,Irish}. One can show \cite{Huang} that introducing a special type of frequency modulations applied to the field and the qubit will give the dynamics governed by Jaynes-Cummings Hamiltonian with $\gamma_f$ in the range of deep strong coupling regime. In circuit-QED simulations rates of $\gamma_f$ for Rabi model up to 2.1 are achieved~\cite{Yoshihara, Langford}. Thus value is lower, while not that much, than the minimal $\gamma_f\approx 7.5$ that is necessary to see the effect in which minimal dimension of the error correcting code in the proposed scheme will be 3 or greater. Our analysis  allows to construct non-trivial quantum error correcting codes for all possible values $\gamma_f,\gamma_s$. 
	
We remark that our scheme could be applied to any system that possess the same decomposition into the direct sum, where eigenenergies in the two direct summands form strictly increasing sequences. Potentially, this property could be exploited for more complex Hamiltonian beyound the Jaynes-Cummings model, as for example for Jaynes-Cummings-Hubbard Hamiltonian, describing interaction of several qubit-cavity systems, or for perhaps directly for Rabi Hamiltonian.

\section{Conclusion}
In this work, the theory of non-commutative operator graphs has been developed for error correction in the case of a finite-dimensional quantum system coupled to an infinite-dimensional quantum system. We have constructed the non-commutative operator graph generated by orbits of the unitary group driven by Hamiltonian (\ref {HAM}) of the Jaynes-Cummings model. We have shown that for a positive integer $K_0$ that satisfies (\ref{nontriv}), using for encoding the eigenstates $\ket {g,0}$ with $\ket {n ,-}$ for $0\le n<K_0$ (\ref {dressedstates}) allows to transmit information with zero error via quantum channels with operator graphs belonging to the constructed graph. Thus the error correcting subspace is explicitly computed for all values of the parameters of the Jaynes-Cummings model. Our scheme could be applied to any system that possess the same decomposition of eigenenergies into the direct sum as for JC Hamiltonian, where eigenenergies in the two direct summands form strictly increasing sequences.

\begin{acknowledgments}
This work was funded by the Ministry of Science and Higher Education of the Russian Federation (grant number 075-15-2020-788) and performed at the Steklov Mathematical Institute of the Russian Academy of Sciences.
\end{acknowledgments}

\end{document}